# A PROPOSED ARCHITECTURE FOR NETWORK FORENSIC SYSTEM IN LARGE-SCALE NETWORKS

Tala Tafazzoli[1], Elham Salahi[2], Hossein Gharaee[3]

[1,3]Network Security Technology Group, Iran Telecommunication Research Center, Tehran, Iran
tafazoli@itrc.ac.ir, gharaee@itrc.ac.ir

[2]IT Security and Systems Group, Iran Telecommunication Research Center, Tehran, Iran

e_salahi@itrc.ac.ir

## ABSTRACT

*Cybercrime is increasing at a faster pace and sometimes causes billions of dollars of business- losses so investigating attackers after commitment is of utmost importance and become one of the main concerns of network managers. Network forensics as the process of Collecting, identifying, extracting and analyzing data and systematically monitoring traffic of network is one of the main requirements in detection and tracking of criminals. In this paper, we propose an architecture for network forensic system. Our proposed architecture consists of five main components: collection and indexing, database management, analysis component, SOC communication component and the database.*

*The main difference between our proposed architecture and other systems is in analysis component. This component is composed of four parts: Analysis and investigation subsystem, Reporting subsystem, Alert and visualization subsystem and the malware analysis subsystem. The most important differentiating factors of the proposed system with existing systems are: clustering and ranking of malware, dynamic analysis of malware, collecting and analysis of network flows and anomalous behaviour analysis.*

.

## 1. INTRODUCTION

By continuing to use technology, cyber-attacks occur rapidly and malware spreads across the globe. Considerable countermeasures have been developed to protect and react to cyber-attacks and cyber-crime.
Despite the reactive and preventive security measures taken to protect networks, forensic investigation of critical information infrastructure is still necessary. Security operation centres (SOC) store security alerts produced by network security appliances. They also store partial network traffic. On the other hand, network forensic systems must store all network traffic. Digital forensic process and investigation depends on the information stored in security operations centre (SOC) communication channel, other network appliances and raw traffic stored in forensic system. The appliances needed to perform the forensic process on this communication channel help the investigators to answer the six questions in forensics: why, how, when, what, where, and who committed the crime.

Digital forensic process for cyber-attack and cyber-crime investigation is an intelligent task. To perform the investigation successfully, the investigator has to be innovative and intelligent. Observing the attack and crime evidence, the investigator intelligently extracts the information from traffic and alerts stored in security appliances. After analysing the evidence, the investigator either accepts or rejects the hypothesis.

In this paper, we present an architecture for network forensic solution, in which we cover weaknesses of current commercial products and research frameworks. Commercial products and solutions establish themselves in analysis capabilities such as session reconstruction, signature analysis, statistical analysis and searching methods with reasonable speed and storage capabilities such as supporting of 1G or 10G traffic. Research frameworks focus on different aspects in network forensics such as distributed storage and searching of evidence, use of soft computing based methods in forensic analysis, etc. Some key components make up our architecture, bringing advanced capabilities and nearly unlimited scalability to bear on network traffic monitoring, analysis problems, anomalous behaviour analysis, online executable filtering, online processing, clustering and analysis of executable to our proposed architecture. The purpose of this paper is to present a unique network forensics tool that will allow the network forensics examiner to participate more effectively in the analysis of a network-crime-based investigation. Using this network forensics tool, the network forensics examiner can enhance the success of solving the case attributable to the accurate, timely, and useful analysis of captured network traffic for crime analysis, investigation, and/or intelligence purposes.

The following chapters are organized as follows. Network forensic architectures and frameworks are presented in chapter 2. Our proposed architecture of network forensic system for large scale networks like telecommunication infrastructure network is presented in chapter 3. We compare our proposed architecture with existing network forensic systems in chapter 4.

## 2. EXISTING FRAMEWORKS AND ARCHITECTURE

The following formatting rules must be followed strictly. This (.doc) document may be used as a template for papers prepared using Microsoft Word. Papers not conforming to these requirements may not be published in the conference proceedings.

Every organization or network requires a specific architecture for its forensic system depending on its characteristics and goals. Various frameworks and architectures for network forensic systems have been introduced [1]. These frameworks can be classified as follows: distributed system frameworks [2, 3, 4, and 5], dynamic network frameworks [6], soft computing based frameworks [7, 8, 9, and 10] and graph based frameworks [11]. Further, we investigate an example framework of each class in detail [5].

Fornet is a distributed network forensic framework. In 2003, Shanmugasundaram et al. introduced Fornet which is a distributed network logging mechanism over wide area networks. This system is developed for digital forensic purposes. The framework is composed of two components: SynApps and Forensic Server. SynApp integrates to network devices, such as switches and routers and summarizes and remembers network events for a time interval and is able to verify these events with certain confidence levels. Although a single SynApp can provide very useful information to a forensic analyst, a network of such co-operating appliances would bring powerful new possibilities to the types of information that could be inferred from the combined SynApps. Networking SynApps would also help to share data and storage and answer to the queries accurately. These SynApps can be organized in a peer-to-peer architecture to collaborate with each other in absence of centralized control, although a hierarchical architecture is simpler and would work better with the given structure of the Internet.

In the hierarchical architecture all the SynApps within a domain form a network. They are associated with the Forensic Server of that domain. In fact, Forensic Server is a centralized administrative control for the domain which manages a group of SynApps in that domain.

Forensic Server receives queries from outside of its domain and processes them with the help of the SynApps and passes the results back to the sender after authentication and certification. Network of SynApps form the first level of hierarchy in ForNet hierarchical architecture. Forensic Servers can also be networked for inter-domain collaboration which forms the second level of the hierarchy. Queries that need to cross domain boundaries go through appropriate Forensic Servers. A Forensic Server is the only gateway to queries sent to a domain from outside the domain boundaries. In other words, a query sent to a domain goes to the Forensic Server of that domain, is authenticated by the server and passed on to appropriate SynApps in the domain. Likewise, the SynApps process results are sent to the Forensic Server that is in control of the domain to be certified and sent back. In practice, queries begin from the leaf nodes of a branch in the hierarchy, traverse Forensic Servers in higher levels, and end up in leaf nodes in another branch. Queries usually travel in the opposite direction of the attack or crime.

In 2007, Wang et al. proposed dynamic network forensic model based on the artificial resistance theory and multi agent theory [6]. The system presents a moment method to collect and store log data simultaneously. Furthermore, the system has the capability to collect the evidence automatically and to respond to network crimes rapidly. Agent theory is a new method for designing, analysing and implementing an open system. A group of agents solve the problems that cannot be handled by each of the agents. These agents have the same role as the protective white cells in human body. Therefore, the agent based security system is more flexible and is similar to a human body immunity system. Dynamic forensic network is used in large scale networks and stores digital data in a distributed and secure manner, when attacks happen.

Soft computing approaches in network forensic systems deal with analyzing collected data and classifying related attacks. Neural network and fuzzy tools are used to verify the occurrence of attacks. In 2007, Zhang et al. proposed a network forensic system based on neural networks and feature extraction with ANN-PCA [10]. Storing and analyzing large amounts of information poses important challenges for network forensic experts. ANN-PCA correlates features with attacks and reduces the amount of stored data. ANN-PCA techniques are used for fraud detection, feature extraction and signature generation for new attacks. FAAR algorithm is used to classify and to search for relational rules and to compute PCA values. Feature extraction in ANN-PCA increases classification accuracy and decreases the data storage amount.

In 2008, Wang and Daniels proposed a graph-based approach for network forensic analysis [3]. The brief description of the main components of this work is as follows:

- Evidence collection module, which collects digital evidence from heterogeneous sensors on networks and hosts.
- Evidence preprocessing module, which transforms collected evidence into standardized format and reduces the redundancy in raw evidence.
- Attack knowledge base, which provides knowledge of attacks, their phases and target vulnerabilities.
- Assets knowledge base, provides knowledge of the networks and hosts under investigation, including network topology, system configuration and value of entities.
- Evidence graph manipulation module, generates the evidence graph by retrieving preprocessed evidence from the depository. Hypotheses and out of band information are also instantiated into the evidence graph through graph edit operations.
- Attack reasoning module, performs semi-automated reasoning based on the evidence graph. In the hierarchical reasoning process, results of local reasoning provide instant updates to the evidence graph.

## 3- A REVIEW ON NETWORK FORENSIC SYSTEMS

NetWitness Company [17] offers modular hardware and software network forensic solutions. Considering the modular and scalable architecture of Netwitness products, they can be used in small sized companies and organizations, large multinational enterprises, datacentres and ISPs. NetWitness architecture consists of five subsystems: network traffic collection subsystem, data processing subsystem, data synchronization subsystem, indexing subsystem and analysis subsystem.

SiLK is an open source analysis software [18] which consists of a set of command-line tools which processes flow records collected through SiLK Packing System. SiLK tools read and process flow records gathered in binary format (segmentation, sorting and analysis). Some of the analysis capabilities of SiLK include: filtering, displaying and sorting, counting and classifying, processing and storing statistical information, labelling based on port number and IP address.

Silent Runner is a tool used for collecting and visualizing the data. Some of the analysis capabilities of this tool are: correlating network traffic with log and alert files, analysis of content and pattern, analysis of requested security incident. This tool has a simple and flexible architecture. Silent Runner monitors and analyzes network and application layer data.

## 4. OUR PROPOSED ARCHITECTURE

Our proposed network forensic system architecture for large-scale networks is presented in Figure 1. The main components of this system are as follows:
- Network traffic collection and indexing subsystem
- Database management subsystem
- Analysis subsystem
- SOC Communication part
- Database.

Each of these components are described in the following subsections.

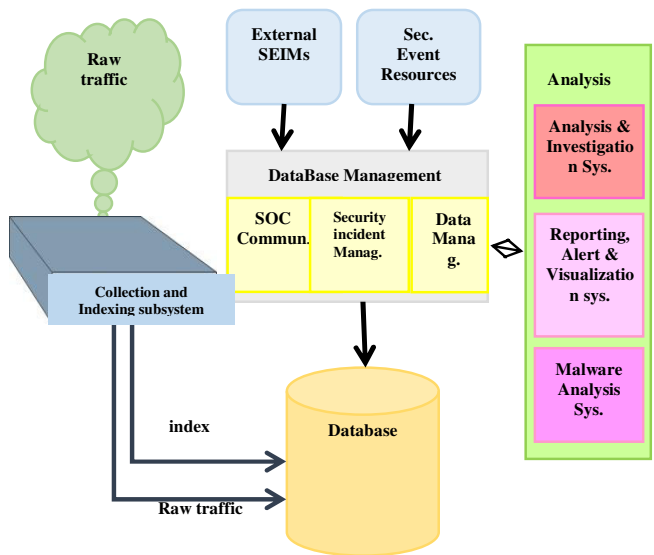

Figure 1 - The Proposed Architecture for our network forensic system

### 4.1. Network Traffic Collection and Indexing subsystem

Traffic collection is referred to the set of the following operations: Receiving traffic through the network interface or importing data from an external file or database, Applying filters or storing the traffic in the system database for further use in next phases. We use indexing to provide quick access to the required data among the mass data. Indexing and the method to perform it define an important part of system speed parameter both in receive and store phase and in analysis phase [5, 6]. Indexing is an important parameter that has implications on network forensic system speed, in both collection and analysis subsystems [5, 6]. The implementation on indexing component depends on the file system. Macro architecture of traffic collection and indexing subsystem is shown in figure 2. We describe it in the following subsections.

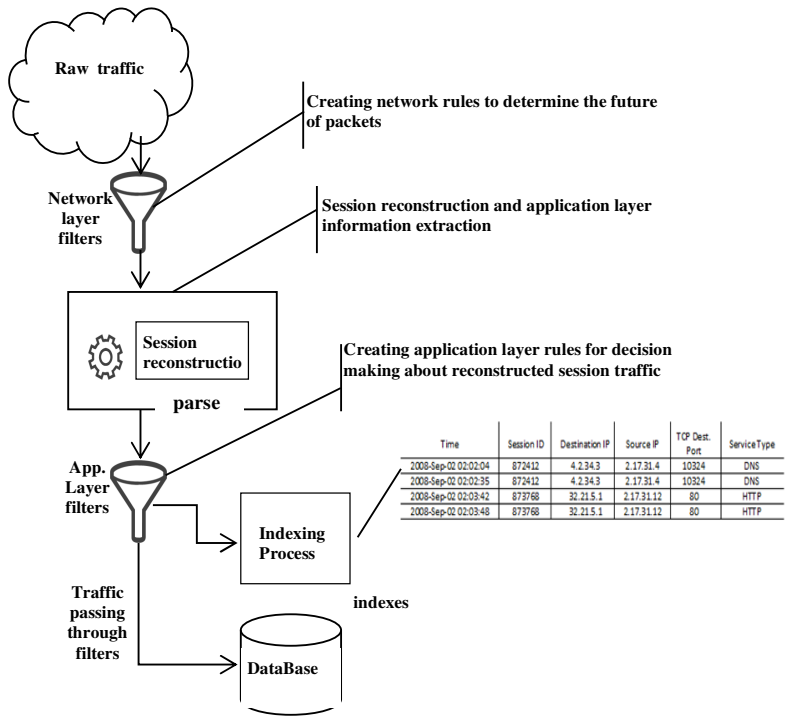

Figure 2 - Macro Architecture of Traffic Collection and Indexing Subsystem

### 4.1.1. Network Traffic Collection

Traffic data can be collected with two different technologies. First, span/mirror ports indicate the ability to copy traffic from all/one port to a single port. Second, Rap devices monitor the traffic flowing between two points in the network.
Both methods have their own advantages and disadvantages, and each of them can be used in a forensic system with proper analysis and consideration of the expected level of advantages and disadvantages. SPAN ports have lower cost and lower reliability compared with TAP devices. Routers doesn't send received packets with inconsistent CRC to SPAN ports. Another disadvantage of SPAN port is the possibility of bandwidth overflow of a port and packet loss. The reason is that the traffic of all the ports are transmitted through one single port. Unlike the SPAN port, TAP devices send a copy of all packets transmitting on an input line to the network to two separate lines with the same bandwidth as the original line. Currently, in SOC systems, data is transmitted to Intrusion Detection Systems using TAP devices. An image of the traffic can also be stored in the forensic system. The policy for data collection is determined by the company using this system. Moreover, our forensic system must have the capability to import and index network traffic data using network packet files. Our system must be compatible with different network traffic file types.

### 4.1.2. Network Filters

Network layer filters separate the packets without session reconstruction. In other words, in this type of filtering, packets are classified based on their headers. Header information of three TCP/IP layers – network, internet and transport layers- is used to separate the traffic without session reconstruction. Some of the header information used for traffic filtering is shown in Table 1.

Table 1 - Protocol Stack Layers and their Characteristics for filtering

| TCP/IP LAYER | PROTOCOL | q |
|---|---|---|
| Network Interface | Ethernet | Address Destination Source Type |
| Internet Layer | IPv4 | Address Destination Source Protocol |
| | IPv6 | |
| Transportation Layer | TCP | Source port Destination Port |
| | UDP | |

### 4.1.3. Parsers
Parser is a program that receives continuous string of symbols and parses it to defined units. Parsers are usually one of the main components of a compiler in programming languages. In this article, parsers are used to scan the information of a reconstructed session and search for certain characteristics in that string. For example, a parser specific for MAIL Standard protocol is used to scan an e-mail session. The parser scans the whole session to find information such as source account, content, etc. The parser put this information in the memory for application layer filtering.

### 4.1.4. Application Layer Filters
After session reconstruction and extraction of application characteristics, the necessary rules for traffic filtering based on application layer characteristics should be developed. Filtering at this layer is the same as lower layers and traffic filtering is performed based on application layer rules. For example, a rule can be defined to store the header information of SSL packets without payload. The goal of network and application layer filters is to limit input/output traffic storage.
In both filters (network and application layer), the system admin should be able to choose the possible actions performed on the packet considering the rule i.e. the future of the packet is determined based on the rule applied to it. Some examples of possible actions on the packet are: session reconstruction of the packet, storing network metadata in the process of data mining. All possible actions for the rules should exist in the system.

Network traffic collection and indexing subsystem resides on an isolated hardware and is placed in network endpoint. It receives a copy of traffic from the Mirror port as input. Scalability and distribution are basic requirements of the collection subsystem. The subsystem must be able to receive network traffic from multiple points in the network simultaneously. For example, suppose that network traffic is received from one 10G Ethernet port and two 1G Ethernet ports. In this case, three traffic collection subsystems are installed in the network separately. The metadata collected by each of these collection subsystems, must be integrated and stored in the system.

### 4.1.5. Indexing
In this phase, indices are generated from network traffic to speed up database access (see figure 2). Indices provide quick access to data. Indexing and the way it is performed, is very important in determining the speed of system in collection and analysis phases.

### 4.2. The Analysis System
The analysis system is used to analyze the stored data in database and consists of three parts: Analysis and Investigation system, Report, Alert and Visualization System and Malware Analysis System.

### 4.2.1. The Analysis and Investigation Subsystem
This system provides analysis capabilities for raw traffic stored in database. Analysis in this system is based on counting (counting the packets, sessions, etc.). Metadata generated in this system consist of different parts (such as source IP address, destination IP address list). The analysis and investigation system provides the capability to focus in metadata, part of metadata, sessions in some part of metadata and to investigate the information related to that particular part more closely. For example, it should be possible to query for metadata for sessions with certain source IP address, related services such as TCP/IP/HTTP, session size in KB and to query for related events and to reconstruct their sessions [7].
Network analyses are performed on layers 2 to 7. Analysis can be done in the following ways: session analysis, path and content analysis, finding the sources of external threats, mapping IP addresses to geographic location, storing data in the sessions, displaying data as seen by the user (web, voice, email, chat, files, etc.), searching and analyzing the contents and sessions (MAC, IP, keyword, usernames, …), the ability to define parsers and alerts, the ability to define customized operations.
Some of the results and outputs of this unit are: alerts, Source and destination IP addresses, source and destination port numbers, IP address, title, email content, session graph display, server program, service type (HTTP, SSL, DNS, …), sites used, attachments, file types, username, reconstructed sessions and protocols restored content such as web page and email content, decrypted and decompressed content.
The following analysis methods are provided by this unit:
*Temporal Analysis:* storage and display of sessions and operations in chronological order and limiting the investigation to a particular range of time.
*Packet Level Analysis:* displaying the number of packets for each metadata.
*Session Level Analysis:* displaying the number of sessions for each metadata.
*File Content Analysis:* Specifying the name, type and origin of metadata.
Moreover, the system should provide the capability to perform these analyses:

*Temporal Analysis:* Displaying the start time and end time of incidents to create incident time axis, displaying incident duration. For example, displaying how fast the malware spreads and the life time of an incident.
*Incident Source Analysis:* the source of the attack is determined.
*Incident Destination Analysis:* the destination of the targeted attack is determined.
*Relation Analysis:* the relations between attack sources and also the relations between culprit and victim is determined.

### 4.2.2. Reporting, Alerting and Visualization Subsystem
This subsystem is connected to various databases. In this subsystem, graphical representations of volume of traffic and protocols used are created. Files, contents and exchanged information with different protocols are extracted and displayed in visualization unit. In this subsystem, the information stored in different databases is investigated and online reports are created.

### 4.2.3. Malware Analysis system
The system architecture is shown in Figure 4. It consists of four main units: file signature control unit, network anomalous behaviour detection unit, automatic sandbox, and domain knowledge unit. Each of these units is described as follows.

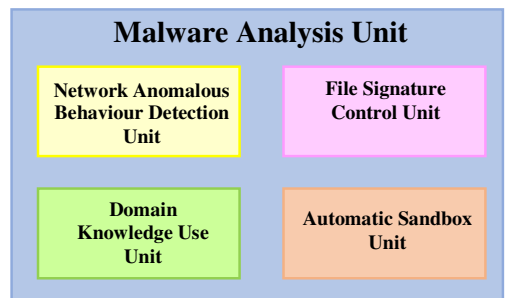

Figure 4- Malware Analysis system components

**File Signature Control Unit:** Malware file signatures are created in this unit and are compared with signatures of executables that are transmitting over the network.
**Network Anomalous Behaviour Detection Unit:** In this unit, with respect to network traffic characteristics, network bandwidth and server capacity thresholds are determined to detect anomalous behaviour in the network. Behavioural profiles are also created. Moreover, traffic is analyzed online and in real time from the network behavioural perspective. In this method, network behavioural pattern in a time period is considered as network's normal behaviour and network normal pattern in time is measured based on it. Using this method, some of network risk areas which are not addressed in other methods are covered. The other analyses are not addressed in this system.
**Domain knowledge unit:** In this unit, if possible, latest malware signatures are received from other companies inside and outside of the company.
**Automated Malware Analysis section:** In this unit, executable files transmitting over the network are first extracted. Unknown executable samples run in sandbox controlled environment and their behaviour are logged. Malware behaviour profile is extracted from recorded logs, network traces, registry changes and access to files. This profile is used for different purposes. Identification and seizure of malwares transmitting over the

network are two main applications of this section. This section is composed of two main units:
- preprocessing and filtering
- Automatic sandbox [11,12]

Figure 5 shows the overall scheme of the unit. Automated malware clustering and analysis have the following benefits:
- A new executable is analyzed rapidly and is determined whether it belongs to known or unknown family.
- If malware belongs to a known family, there would be no need for dynamic analysis. Therefore fewer samples are analyzed in sandbox. The sandbox platform should be capable of online analysis of malware samples.

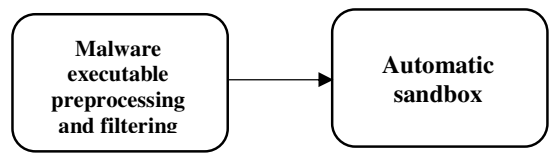

Figure 5- Overall Scheme of the Automatic Malware Analysis Unit.

The main responsibilities for each of the units of this section will be explained sequentially.

**Preprocessing and Filtering Unit:** The purpose of this unit is principally focused on extracting executable codes, reducing their count and separating unknown samples. There are different approaches to achieve this goal. We explain one approach here. This approach is shown in figure 6.

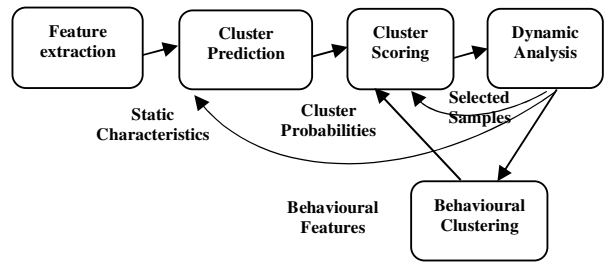

Figure 6 - A Scheme for Executable Code Preprocessing and Filtering Unit

Here we explain different units of this scheme [8, 9, and 10].

**Feature extraction.** First, static features of each sample are extracted without executing the code. They are extracted from the code. Some of these features are as follows: file structure, antivirus analysis results of the code.

**Cluster prediction.** Behavioural profiles and static features of the code are used as input to this unit. Here, we attempt to predict the behavioural cluster of the sample, using a supervised learning approach.

**Cluster scoring.** In this phase, a score is assigned to each sample which investigates the dissimilarity of the sample in the cluster. High score samples are passed to dynamic analysis phase.

**Dynamic analysis.** Selected executable samples are executed in the sandbox. The network level and host level behaviour observed during execution is condensed in a set of behavioural features. The set of features is fed back to the cluster ranking phase. We cluster malware samples based on their malicious behavioural features. The behavioural cluster that a sample belongs to, is fed back to cluster prediction phase. Therefore, high score samples from preprocessing phase are fed as input to dynamic analysis phase and behavioural characteristics of malware samples are output from the sandbox. These characteristics are stored in a behavioural profile. The profiles are expressed by OS objects and OS functions.

### 4.3. Database Management
Security Operation Center is responsible for receiving network security events from different security tools, and correlating and analyzing them.

Connection to Security Operation Center occurs in data management section. A general view of connection to security operation center is shown in figure 7. Network forensic system searches for patterns. The requests are sent to security operation center and appropriate responses are received from database management system in an acceptable time.

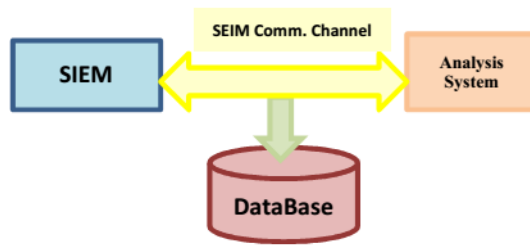

Figure 7- Communication to the Security Operation Center

APIs should be placed between SOC and analysis system to covert data structures transmitting between different units. Database management unit should combine raw traffic with alerts transmitted from security operations center. Considering Telecommunication Infrastructure policies, the forensic system might be authorized to access and search in alert databases and correlation logs directly. Searches must be performed with proper speed.

### 4.4. Database
There are two methods for storing traffic and metadata. The first method is clustered storage [16]. In this storage, all raw traffic and metadata are aggregated and stored in server farm. The advantages of this technology include: integration of traffic stored in the network to be used by all security units, reducing setup and maintenance costs, and increasing the system security. In contrast, implementation complexity and the need for information about system implementation details are possible disadvantages of clustered storage subsystem.

### 5. A comparison between our method and other proposed method
Table 2 shows different aspects and components of NetWitness, Silk and Solent Runner system and compares them to our proposed architecture. It shows strengths and weaknesses of our architecture in comparison with other products.

# 6. Conclusion

The main purpose of forensic systems is to help the investigators to identify criminals and crime signs. Strength of a network forensic system depends on its ability to process network traffic with a speed proportional to data transmission and analysis.

In this paper, we propose an architecture of a forensic system for security operation center. It is composed of several units: data collection and indexing, database management system and analysis system. The proposed architecture tries to use maximum existing potentials to optimize designed system efficiency. Compared to existing frameworks, our proposed architecture brings us advanced capabilities for dynamic malware analysis and clustering and analysis and network behavioural analysis.

Table 2- The proposed architecture compared with existing systems

| Product Name / Capabilit | Proposed Architecture | Silent Runner | Silk | Netwitness NextGen |
|---|---|---|---|---|
| **Data and Traffic Collection Characteristics** | | | | |
| 1G and 10 G Traffic Support | ✓ | ✓ | ✗ | ✓ |
| Wireless Network Sniffing Capability | ✗ | ✓ | ✗ | ✓ |
| VoIP Call Recording (CDR) | ✓ | ✓ | ✗ | ✓ |
| **Analysis Capability** | | | | |
| Session Restore | ✓ | ✓ | ✗ | ✓ |
| Network Behaviour Analysis | ✓ | ✓ | ✗ | ✓ |
| Statistical Analysis | ✓ | ✓ | ✓ | ✓ |
| Anomalous Behaviour Analysis | ✓ | ✗ | ✗ | ✗ |
| Malware online Detection | ✓ | ✓ | ✗ | ✓ |
| Online Dynamic Code Analysis | ✓ | ✗ | ✗ | ✗ |
| Malware Clustering | ✓ | ✗ | ✗ | ✗ |
| Executive Code Filtering | ✓ | ✗ | ✗ | ✗ |
| Executive Code Filtering and Processing | ✓ | ✗ | ✗ | ✗ |
| Acceptable Search speed in Raw Traffic | ✓ | ✓ | ✗ | ✓ |
| Protocol Restore (at least 5 Protocols) | ✓ | ✗ | ✗ | ✓ |
| Network Flow based Analysis | ✓ | ✗ | ✓ | ✗ |
| Raw Traffic Indexing | ✓ | ✓ | ✗ | ✓ |
| SSL/HTTPS key Exchange Proxy | ✗ | ✗ | ✗ | ✗ |
| Alert Notification System | ✓ | ✓ | ✓ | ✓ |
| **Architecture Characteristics** | | | | |
| Scalability | ✓ | ✓ | ✗ | ✓ |
| Distributive Capability | ✓ | ✓ | ✗ | ✓ |
| Development Capability | ✓ | ✓ | ✓ | ✓ |
| Present Device with P&P Capability | ✓ | ✓ | ✗ | ✓ |
| Remote Search Capability | ✓ | ✓ | ✗ | ✓ |
| Open Source | ✗ | ✗ | ✗ | ✗ |

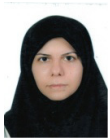
**Tala Tafazzoli** received B.S. degree in computer engineering from Sharif University of Technology and received M.S. degree in computer engineering from Amirkabir University of Technology. She is a faculty member of network security group of Iran Telecom Research Center. Her research interest areas include: digital forensics, intrusion detection system and network security.

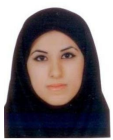
**Elham Salahi** has a BSc degree in Computer Science from Sharif University, and an MSc degree in Applied Mathematics with a minor in Computer science from Iran University of Science and Technology. She has more than nine years of experience working as a researcher in ITRC. Her research interests are: digital forensics, Image Watermarking, intrusion detection system and network security.

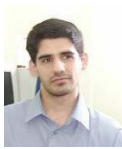
**Hossein Gharaee** received B.S. degree in electrical engineering from Khaje Nasir Toosi University, in 1998, M.S. and Ph.D. degree in electrical engineering from Tarbiat Modares University, Tehran, Iran, in 2000 and 2009 respectively. Since 2009, he has been with the Department of Network Technology in IRAN Telecom Research Center (ITRC). His research interests include general area of VLSI with emphasis on basic logic circuits for low-voltage low-power applications, DSP Algorithm, crypto chip and Intrusion detection and prevention systems.